\documentclass{article}

\usepackage[margin=1in]{geometry}
\usepackage{graphicx}
\usepackage{amsfonts}
\usepackage{amsmath}
\usepackage{authblk}
\usepackage{tikz}\usetikzlibrary{decorations.pathmorphing}
\usepackage{natbib}
\title{Estimating individual contributions to team success\\ in women's college volleyball}
\author[1]{Scott Powers}
\author[2]{Luke Stancil}
\author[3]{Naomi Consiglio}
\affil[1]{Department of Sport Management, Rice University}
\affil[2]{Department of Economics, Rice University}
\affil[3]{Department of Statistics, Rice University}

\begin{document}

  \maketitle
	
  \begin{abstract}
    The progression of a single point in volleyball starts with a serve and then alternates between teams, each team allowed up to three contacts with the ball. Using charted data from the 2022 NCAA Division I women's volleyball season (4,147 matches, 600,000+ points, more than 5 million recorded contacts), we model the progression of a point as a Markov chain with the state space defined by the sequence of contacts in the current volley. We estimate the probability of each team winning the point, which changes on each contact. We attribute changes in point probability to the player(s) responsible for each contact, facilitating measurement of performance on the point scale for different skills. Traditional volleyball statistics do not allow apples-to-apples comparisons across skills, and they do not measure the impact of the performances on team success. For adversarial contacts (serve/receive and attack/block/dig), we estimate a hierarchical linear model for the outcome, with random effects for the players involved; and we adjust performance for strength of schedule not only on the conference/team level but on the individual player level. We can use the results to answer practical questions for volleyball coaches.
  \end{abstract}

\section{Introduction}
\label{sec:intro}

Women's volleyball has recently grown in popularity in the United States. In 2023, the University of Nebraska-Lincoln hosted an open-air volleyball match against the University of Nebraska-Omaha at Memorial Stadium, typically used for football games. At 92,003, the event set a world record for attendance at a women's sporting event \citep{olson_2023}. In 2024, the Pro Volleyball Federation debuted its inaugural season, with another professional league---League One Volleyball---expected to debut later in the year \citep{echlin_2024}. Of note for the present work, volleyball lends itself well to statistical analysis.

A volleyball match has the following structure: Two teams of six are positioned at opposite sides of a net that bisects the court horizontally. For statistical purposes, each half of the court is further divided into 9 zones as shown in Figure \ref{fig:volleyball-court-diagram}. The first team to win 3 {\it sets} wins the game. The first four sets are played to 25 points, win by 2, meaning that a set is played until a team has at least 25 points and a lead of at least 2. The fifth set, if needed, is played to 15 points, win by 2. Teams score points by winning rallies. A rally is won in if: the ball touches the ground in the playing area on the opponent's side of the court, the ball lands outside of the playing area last touched by the opponent, or the opponent commits a violation (such as contacting the net). Each rally has three distinct phases of play: service, first ball, and transition. A rally is initiated with a serve (service) that must go over the net to the other team's side. When a team receives the ball from the serve, it begins the first-ball phase and can contact the ball a total of three times before the ball must cross the other side or result in a violation. After the ball crosses the net, the transition phase begins and the other team now must return the ball within three contacts (block contacts do not count toward this limit). We use the word {\it possession} (some use ``complex'') to describe the sequence of consecutive contacts that a team performs while the ball is on their side.

Among the six starting players on the court, most teams deploy players in four different positions. The setter (S) is primarily responsible for setting the ball to the attackers. The two outside hitters (OH) are primarily responsible for attacking and passing. The two middle blockers (MB) are primarily responsible for blocking and attacking. The opposite hitter (OPP) is primarily responsible for attacking and blocking. In addition to these roles, there is a special position, the libero (L), who wears a different-colored jersey. The libero can freely come on and off the court for any player, but she can only play in the back row, so she is primarily responsible for passing. Every time a team wins service back from their opponent (a team always serves after winning point), the players rotate one position clockwise around the court. Although players are required to start the point in this rotation position, generally they will quickly run to their desired position on the court as soon as the ball has been served (subject to the constraint that back row players cannot play front row positions).

The substitution tendencies of a team depend on the type of offense that they run. Generally, one of two offenses is used by a team (and may be switched in the middle of a match): 6-2 or 5-1. A 6-2 refers to having 6 hitters and 2 setters throughout the rotation, and a 5-1 refers to having 5 hitters and 1 setter throughout the course of a rotation. When a team runs a 6-2, they will always have their setter in the back row. When a setter rotates into the front row, they will be replaced by an opposite hitter. This means that the team will always have three attackers in the front row and will have zero or one possible attacker in the back row. Conversely, a 5-1 will see the setter remaining the game to play front row.

The standard box score metrics for the core skills in volleyball are ace percentage and error percentage for serving; error percentage and passer rating (ordinal rating of pass quality) for receiving; digs per set and digs per opportunity for digging; assits per set for setting; hitting efficiency for attacking, defined as (Kills -- Errors) / Attempts; and blocks per set for blocking. All of these measurements are in different units, and their is no obvious way to compare performance across skills.

Our approach is to implement two techniques to measure individual contribution to team success in volleyball. First, we will use a Markov chain to estimate the probability of each team winning the point, updating the probability as each contact occurs. We will aggregate changes in win probability to measure those performance in units of points. This will facilitate comparisons across skills. Second, we will use a hierarchical regression model (random-effects linear regression) to estimate strength of schedule on a contact-by-contact basis. This will facilitate comparisons across conference that have disparate levels of competition.

\subsection{Related Work}
\label{sec:related-work}

The Markov chain has long been used to model the progression of sports which have well-defined states and probabilistic transitions between them. Baseball \citep{bukiet_etal_1997} may be the sport where this type of analysis is most predominant, with the game state defined by which bases are occupied and how many outs there are. The model has also been used in tennis \citep{newton_etal_2009} and table tennis \citep{pfeiffer_etal_2010}, which are net sports like volleyball.

Two early uses of the Markov chain model for skill measurement in volleyball are \citet{florence_etal_2008} and \citet{miskin_etal_2010}, both of which used contact-by-contact data (contacts by the home team only) from one season of home matches for a college women's volleyball team. More recently, \citet{bagley_ware_2017} used a full season (over 4.5 million contacts) of contact-by-contact data from women's college volleyball to rate player performance in the core skills.

\citet{drikos_2018} used a Markov chain model with international men's (and youth) volleyball data to estimate the probability of winning a point based a rating of pass quality, with the goal of validating a six-level ordinal scale for pass quality. \citet{hileno_etal_2020} used a Markov chain model with international women's volleyball data for the sequencing of categorized possessions, with the goal of identifying possession categories for focused training.

Hierarchical regression modeling has also been used recently in volleyball. \citet{gabrio_2021} used several different Bayesian hierarchical regressions to model the results of women's professional volleyball matches in Italy. \citet{fellingham_2022} used Bayesian hierarchical logistic regression to model the outcomes of possessions based on which players contact the ball during the possession. This method measures individual contributions to team success across varied skills on the scale of points, so it is the most similar to the present work.

What distinguishes the present work is that no prior work measures the change in point win probability with each contact and aggregates those changes to evaluate each skill as it impacts team win probability. Additionally, no prior work adjusts player performance measurements based on the quality of competition faced by the player. These two contributions are important for informing decision-making by college women's volleyball teams, both in player recruitment and in-game strategy.

\section{Data}
\label{sec:data}

For every contact that occurred in the 2022 NCAA Division I women's college volleyball season, our dataset includes the identity of the player; the type of skill; the start and end locations of the event; an evaluation of the quality of the contact; and more. To illustrate the date, Table \ref{tab:sample-data} shows some of the columns, using the first point of the 2022 national championship as an example.

\begin{table}[hb]
    \centering
    \begin{tabular}{cccccccc}
   Possession   & Player          & Team         & Skill                 & Eval          & (X, Y)                  & Attack Code  & End Zone\\
  \hline
   1            & Anna Deeber         & Louisville &  Serve         &  --    &  (2.99, -0.13) \\
   2            & Emma Halter         & Texas &  Reception     &  \# &  (0.93, 5.80) \\
   2            & Saige K.-Torres     & Texas &  Set           &  \# &  (2.13, 3.13) \\
   2            & Molly Phillips      & Texas &  Attack        &  -- &  (3.33, 3.20) &  X6             & 9\\
   3            & Raquel Lazaro       & Louisville &  Dig           &  +     &  (0.86, 4.98) \\
   3            & Elena Scott         & Louisville &  Set           &  \#    &  (2.99, 1.65) \\
   3            & Claire Chaussee     & Louisville &  Attack        &  --    &  (0.63, 2.83)    &  V5  & 9\\
   4            & Kayla Caffey        & Texas &  Block         &  +  &  (3.26, 3.43) \\
   5            & Phekran Kong        & Louisville &  Dig           &  !     &  (0.89, 3.13) \\
   5            & Raquel Lazaro       & Louisville &  Set           &  \#    &  (0.97, 2.61) \\
   5            & Claire Chaussee     & Louisville &  Attack        &  \#    &  (0.67, 2.91)    &  X5  & 5\\
\end{tabular}
    \caption{\it An illustrative sample from the dataset. These are the contacts from the first point of the 2022 NCAA national championship, with the most important columns shown. For each contact, we observe the player, the skill, an evaluation of the quality of the contact, and the coordinates of the starting location. For attacks, we also observe the attack code describe what type of attack it is.}
    \label{tab:sample-data}
\end{table}

\begin{figure}[h]
    \centering
    \begin{tikzpicture}
        \draw[line width = 2pt] (-0.5, 3) -- (3.5, 3);
        \draw (0, 0) -- (0, 6);
        \draw (3, 0) -- (3, 6);
        \draw (0, 0) -- (3, 0);
        \draw (0, 2) -- (3, 2);
        \draw (0, 4) -- (3, 4);
        \draw (0, 6) -- (3, 6);
        \draw[dashed] (1, 3) -- (1, 0);
        \draw[dashed] (2, 3) -- (2, 0);
        \draw[dashed] (0, 1) -- (3, 1);
        \node (1) at (2.5, 0.5) {\bf 1};
        \node (2) at (2.5, 2.5) {\bf 2};
        \node (3) at (1.5, 2.5) {\bf 3};
        \node (4) at (0.5, 2.5) {\bf 4};
        \node (5) at (0.5, 0.5) {\bf 5};
        \node (6) at (1.5, 0.5) {\bf 6};
        \node (7) at (0.5, 1.5) {\bf 7};
        \node (8) at (1.5, 1.5) {\bf 8};
        \node (9) at (2.5, 1.5) {\bf 9};
    \end{tikzpicture}
    \caption{\it Diagram of volleyball court with zones labelled. Within each half of the court, we have nine labelled zones. In the data, these zones are used to describe the starting location and ending location of each contact.}
    \label{fig:volleyball-court-diagram}
\end{figure}
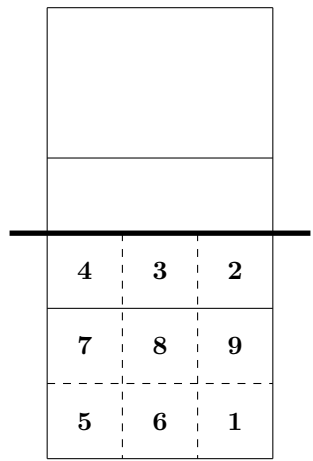

Aside from player, team, skill and lineup information, the most import columns for the present work are the evaluation code, the attack code and the end zone. The evaluation code is detailed in Table \ref{tab:evaluation_codes}. It gives a rating of the quality of the contact. For attacks, the attack code describes the type of attack (set location, set tempo, attacker position), and the end zone indicates the zone of the court where the attack ended, according to Figure \ref{fig:volleyball-court-diagram}.

\begin{table}
    \centering
    \begin{tabular}{ccl}
    Scale & Code          & Description\\
    \hline
    4   & \#            & action results in a point for own team or in the case of a reception or set, is perfect.\\
    3   & +             & action results in opposing team being ``out-of-system'' or own team being ``in-system''\\
    2   & !             & only defined for receptions and is between a ``good'' and ``bad'' reception\\
    1   & --            & action results in opposing team being ``in-system'' or own team being ``out-of-system''\\
    0   & = or \slash   & action is an error and results in a point for the opposing team\\
\end{tabular}
    \caption{\it Five-point evaluation scale and corresponding codes. Every contact is evaluated using this scale, reflecting its quality. Higher values on the scale reflect better performance by the player.}
    \label{tab:evaluation_codes}
\end{table}

The lineup data indicate which player is in each spot in the service rotation at the beginning of each point, and one player is labelled as the setter. This label follows the service rotation regardless of whether the setter subs out of the game (as in a 6-2 offense). Substitutions are recorded in the data, but when the libero comes on and off the court, it is not recorded. In fact, the identity of the libero is not in the lineup data, but we can infer it based on which player is making plays without being in the lineup.

The dataset spans 4,147 matches from the 2022 NCAA Division I women's volleyball season, comprising over 600,000 points and more than 5 million recorded contacts.

\section{Methods}

\subsection{Point Win Probability (PWP) Model}

We model the progression of a volleyball point as a Markov chain, where the state is given by $s = (T, P) \in \mathcal{S}$. $T$ encodes whether the serving (S) or receiving (R) is currently in possession of the ball, and $P$ encodes the subsequence of touches that have occurred in the current possession, including the evaluation code of all touches other than attacks and blocks. For attacks, $P$ also encodes the attack code, i.e. the type of attack. Two terminal states, $(\mbox{S, W})$ and $(\mbox{T, W})$, indicate that the serving team or the receiving team, respectively, has won the point. We use $\mathcal{S}_{\mbox{\scriptsize S}} \subset \mathcal{S}$ to denote the subset of states for which the serving team is in possession, i.e. $T = \mbox{S}$, and $\mathcal{S}_{\mbox{\scriptsize R}} = \mathcal{S} - \mathcal{S}_{\mbox{\scriptsize S}}$ is the subset of states for which the receiving team is in possession, i.e. $T = \mbox{R}$. Every touch corresponds to a transition from a starting state to an ending state.

We exclude the evaluation code for attacks and blocks because these evaluation codes have a very strong one-to-many correspondence with the evaluation code of the following contact. In other words, the evaluation code of the following contact is a strictly richer representation of the outcome of the attack/block, so we rely on this and ignore the evaluation code of the attack/block. One may think of the Markov state ending in an attack as the state of the point when the ball has been set to the attacker, before the outcome of the attack. The next state transition reflects the outcome of the attack.

Using the example data from the first point of the 2022 national championship in Section \ref{sec:data}, the progression of observed states is as follows:
\begin{align*}
    &\mbox{(S, SV)}  \rightarrow\\
    &\mbox{(R, R\#)} \rightarrow \mbox{(R, R\#S\#)} \rightarrow \mbox{(R, R\#S\#AX6)} \rightarrow\\
    &\mbox{(S, D+)}  \rightarrow \mbox{(S, D+S\#)}  \rightarrow \mbox{(S, D+S\#AV5)}  \rightarrow\\
    &\mbox{(R, B+)}  \rightarrow\\
    &\mbox{(S, D!)}  \rightarrow \mbox{(S, D!S\#)}  \rightarrow \mbox{(S, D!S\#AX5)}  \rightarrow\\
    &\mbox{(S, W)}
\end{align*}

For each of state, we want to know the probability of each team winning the point. This would yield a point win probability model that updates with every touch. To estimate the transition probabilities between states, we use the empirical distribution of ending states from each starting state. From this matrix of one-step transition probabilities, we can calculate the $n$-step transition probability for any number of steps $n$ through repeated matrix multiplication. We calculate the 100-step transition probability matrix, which is a sufficiently large number of steps that the end state probability is entirely concentrated in the two terminal states (because the probability of a point lasting longer than 100 touches is close enough to zero for our level of precision). These transition probabilities from each starting state are the estimated point win probabilities from each state.

We use the function $v: \mathcal{S} \rightarrow [0, 1]$ to denote the {\it sideout} probability of each state, i.e. the probability that the receiving team eventually wins the point, reaching the terminal state (R, W). It is also useful to have the point win probability from the perspective of the team in possession, and we use the function $w(\cdot)$ for this.
\begin{equation*}
    w(s) = \begin{cases}
        v(s)        & \mbox{if } s \in \mathcal{S}_{\mbox{\scriptsize R}}\\
        1 - v(s)    & \mbox{if } s \in \mathcal{S}_{\mbox{\scriptsize S}}
    \end{cases}.
\end{equation*}

For further context, we considered richer representations for the point state based on the current rotation of each team. Intuitively, one may expect that the win probability would depend on whether the setter is in the front row or the back row (because a back-row setter has more front-row attacker options). However, we found that this additional information did not lead to a practically significant change in the win probability estimates. Our theorized explanation is that the relationship between rotation and win probability is confounded by individual player talent. For example, teams generally line up their strongest outside hitter and middle blocker next to their setter in the rotation, meaning that having a front-row setter is correlated with having more talented attackers in the front row.

\subsection{Strength of Schedule (SoS) Model}
\label{sec:strength-of-schedule}

With a large number of teams and a wide spread of talent within our dataset, it is no surprise that many of the top individual performances will come from players in smaller conferences that face less stringent competition. From the perspective of player scouting and recruitment, it is helpful to adjust individual performances according to their strength of schedule, i.e. the quality of competition they faced. Generally, our approach is to model the outcome of adversarial touches using generalized linear models (GLMs), with hierarchical random effects for conference, team and player. Because of this hierarchy of random effects, the model regresses team effects toward conference means and regresses player effects toward team means. The two groupings of adversarial outcomes are serve/receive and attack/block/dig. Each grouping yields both a categorical (ending state) and a numeric (win probability from ending state) with each observation. We use different likelihood models for the two groupings.

\subsubsection{Serve/Receive Outcomes}

The serve/receive outcome is the simpler of the two adversarial groupings. The seven possible outcomes are a service error or a dig with one of the following evaluation codes: \# (``perfect pass''), $+$ (``positive, attack''), $!$ (``ok, no first temp possible''), / (``poor, no attack''), $-$ (``negative, limited attack''), = (``error''). That final evaluation code, =, correspond to an ace. Even when no receiver touches the serve, our dataset always includes an assignment of responsibility for the ace to a player on the receiving team. Service errors do not include a receiver responsibility assignment.

We exclude service errors from the SoS model because we do not have receiver responsibility assignments and because we judge the impact of receiver on service errors to be small (using domain knowledge). Therefore our SoS model for serve/receive outcomes is based on the conditional outcome of the serve, given that there is no service error. Rather than modeling the categorical dig outcome, we model the numeric change in win probability corresponding to the dig outcome, using a normal likelihood. Although the normal distribution assumption is unreasonable for a random variable that can only take six different values, the model is fit for purpose because we are only interested in point estimates for the additive effects of conference, player and team on the outcome.

Our dataset contains $n$ non-error serves indexed by $i \in \{1, ..., n\}$. Use use $\mathcal{P}$, $\mathcal{T}$, $\mathcal{C}$ to denote the sets of all players, teams and conferences, respectively, present in our data. For each serve $i$, we observe the server $p_i \in \mathcal{P}$; the serving team $t_i \in \mathcal{T}$; the conference $c_i \in \mathcal{C}$ of the serving team; the receiver $r_i \in \mathcal{P}$; the receiving team $\tilde t_i \in \mathcal{T}$; and the conference $\tilde c_i \in \mathcal{C}$ of the receiving team. Additionally we observe the states $s_i, s'_i \in \mathcal{S}$ before and after the serve outcome, respectively. The pre-outcome state is always $s_i =$ (S, SV). An example of a post-outcome state is $s'_i = (R, R\#)$. We use $y_i = -(\hat v(s'_i) - \hat v(s_i))$ to denote the change in point win probability (from the serving team's perspective) due to the serve outcome. Note that we negate the change in $\hat v(\cdot)$ to frame the win probability gain from the serving team's perspective because $\hat v(\cdot)$ is the sideout probability (i.e. win probability from the receiving team's perspective). We use $Y_i$ to denote the random variable from which the observed data $y_i$ is a random draw. We model the probability distribution of $Y_i$ as:
\begin{align*}
  Y_i &\sim \mbox{Normal}(\eta_i, \sigma^2_\epsilon)\\
  \eta_i &= \alpha + \gamma_{c_i} + \tau_{t_i} + \pi_{p_i} + \tilde\gamma_{\tilde c_i} + \tilde\tau_{\tilde t_i} + \rho_{r_i}.
\end{align*}
The parameters $\sigma^2_\epsilon$ and $\alpha$ are assumed to be fixed, unknown parameters. We model the other components contributing to $\eta_i$ as random effects according to the following specification:
\begin{align*}
  \gamma_c &\sim \mbox{Normal}\left(0, \sigma^2_\gamma\right) \hspace{2mm} \forall\, c \in \mathcal{C} &
  \tilde\gamma_c &\sim \mbox{Normal}\left(0, \sigma^2_{\tilde\gamma}\right) \hspace{2mm} \forall\, c \in \mathcal{C} \\
  \tau_t &\sim \mbox{Normal}\left(0, \sigma^2_\tau\right) \hspace{2mm} \forall\, t \in \mathcal{T} &
  \tilde\tau_t &\sim \mbox{Normal}\left(0, \sigma^2_{\tilde\tau}\right) \hspace{2mm} \forall\, t \in \mathcal{T} \\
  \pi_p &\sim \mbox{Normal}\left(0, \sigma^2_\pi\right) \hspace{2mm} \forall\, p \in \mathcal{P} &
  \rho_r &\sim \mbox{Normal}\left(0, \sigma^2_\rho\right) \hspace{2mm} \forall\, r \in \mathcal{P},
\end{align*}
where all variance components are fixed, unknown parameters. This is a random-effect linear model, and note the nested hierarchy of the effects. Each player belongs to a team, and each team belongs to a conference.

We use the lme4 package \citep{bates_etal_2015} in R to estimate this model, as well as the rest of the random-effect regression models in the present work. Given the random effect predictions resulting from this model, the estimated strength of server $p_i$ (and the estimated strength of schedule for receiver $r_i$) is $\hat\gamma_{c_i} + \hat\tau_{t_i} + \hat\pi_{p_i}$. Some players have small samples, but because of the nested hierarchy of the random effects, the player strength is shrunken toward the team average, and the team average is shrunken toward the conference average.

\subsubsection{Set/Attack/Block/Dig Outcomes}
\label{sec:sos-attack}

We group the outcomes of attacks into five categories: (1) attack error, (2) block error, (3) block-return, (4) block-through and (5) clean attack. An {\bf attack error} occurs when the attacker fails to direct the ball into the opponent's court. For example, the attacker may hit the ball into the net, hit the ball out of bounds or commit a net violation by making physical contact with the net. A {\bf block error} occurs when the blocker does something to immediately lose the point. For example, the ball may bounce off the blocker's hands in a way that is unplayable for her teammates, or the blocker may commit a net violation. A {\bf block-return} occurs when the ball strikes the blocker's hands and returns to the attacker's side of the net. A {\bf block-through} occurs when the ball strikes the blocker's hands and continues on to the blocker's side of the net. A {\bf clean attack} occurs when there is no attacking error and no block touch.

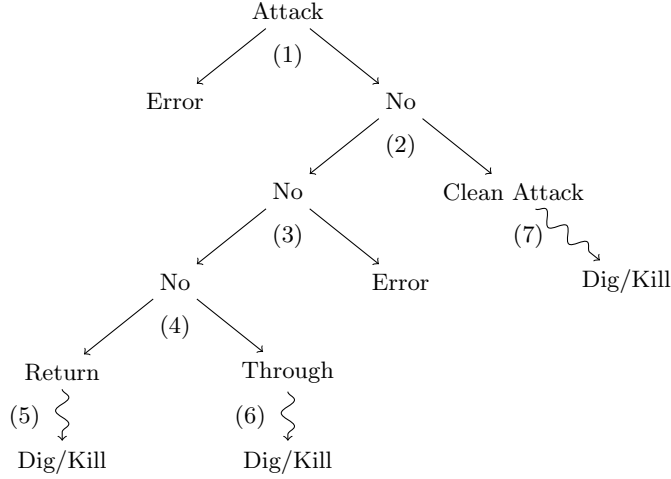
\begin{figure}
    \centering
    \begin{tikzpicture}
        \node (Attack) at (0, 0) {\small Attack};
        \node (Attack Error) at (-1.5, -1.2) {\small Error};
        \node (Attack No Error) at (1.5, -1.2) {\small No};
        \node (Block) at (0, -2.4) {\small No};
        \node (No Block) at (3, -2.4) {\small Clean Attack};
        \node (Block No Error) at (-1.5, -3.6) {\small No};
        \node (Block Error) at (1.5, -3.6) {\small Error};
        \node (Return) at (-3, -4.8) {\small Return};
        \node (Through) at (0, -4.8) {\small Through};
        \node (No Block Dig/Kill) at (4.5, -3.6) {\small Dig/Kill};
        \node (Return Dig/Kill) at (-3, -6) {\small Dig/Kill};
        \node (Through Dig/Kill) at (0, -6) {\small Dig/Kill};
        \draw[->] (Attack) -- (Attack Error);
        \draw[->] (Attack) -- (Attack No Error);
        \draw[->] (Attack No Error) -- (Block);
        \draw[->] (Attack No Error) -- (No Block);
        \draw[->] (Block) -- (Block No Error);
        \draw[->] (Block) -- (Block Error);
        \draw[->] (Block No Error) -- (Return);
        \draw[->] (Block No Error) -- (Through);
        \draw[->, decorate, decoration = snake] (No Block) -- (No Block Dig/Kill);
        \draw[->, decorate, decoration = snake] (Return) -- (Return Dig/Kill);
        \draw[->, decorate, decoration = snake] (Through) -- (Through Dig/Kill);
        \node (1.) at (0, -0.6) {\small (1)};
        \node (2.) at (1.5, -1.8) {\small (2)};
        \node (3.) at (0, -3) {\small (3)};
        \node (4.) at (-1.5, -4.2) {\small (4)};
        \node (5.) at (-3.5, -5.4) {\small (5)};
        \node (6.) at (-0.5, -5.4) {\small (6)};
        \node (7.) at (3.2, -3) {\small (7)};
    \end{tikzpicture}
    \caption{\it An illustration of the attack outcome tree. There are five categories of attack outcomes. At each split in the tree, we estimate a random-effect linear model for the change in point win probability, conditional on the outcome of the split. Each split is labelled according to the equation number for the model describing the split. Straight arrows represent binary splits while wavy arrows represent splits into multiple outcomes.}
    \label{fig:attack-model-tree}
\end{figure}

The attack error and block error have only one outcome: The team committing the error loses the point. The other three categories of outcomes can result in the point ending (if the ball touches the floor) or in a dig, if a player is able to reach the ball before it touches the floor. We use an outcome tree as a taxonomy to represent the hierarchy of attack outcomes, as illustrated in Figure \ref{fig:attack-model-tree}. At the top level of the hierarchy is the split between attack error and not. If there is no attack error, the next split is on whether there is a block touch. If there is a block touch, the next split is between block error and not. Conditioned on a block touch, the final binary split is on whether the ball is returned by the block or gets through the block. For the non-terminal leaves (clean attack, block-return and block-through), we have an ordinal outcome (kill or a dig with some evaluation code).

Our dataset contains $n$ attacks indexed by $i \in \{1, ..., n\}$. As before, we use $\mathcal{P}, \mathcal{T}, \mathcal{C}$ to denote the sets of all players, teams and conferences, respectively, in our data. For each attack $i$, we observe the attacker $a_i \in \mathcal{P}$; the setter $s_i \in \mathcal{P}$; the attacking team $t_i \in \mathcal{T}$; and the conference $c_i \in \mathcal{C}$ of the attacking team. On the other side of the net, we observe the defending team $\tilde t_i \in \mathcal{T}$ and the conference $\tilde c_i \in \mathcal C$ of the defending team. Depending on the outcome, we sometimes observe the blocker $b_i \in \mathcal{P}$, and we sometimes observe the digger $d_i \in \mathcal{P}$. For some outcomes, we do not need to know the blocker or digger. For other outcomes, we infer the block/dig responsibility based on the attack (see Section \ref{sec:block-dig-responsibility}). In addition to the actors involved, we observe the states $s_i, s'_i \in \mathcal{S}$ before and after the attack outcome, respectively. An example pre-outcome state is (R, R\#S\#A), meaning that the attack follows a perfect set which followed a perfect pass in reception. An example post-outcome state is (S, D+), meaning that the outcome of the attack was a good dig for the defensive team.

Our approach is to model the change in point win probability that occurs at each split in the outcome tree. With each split, as we learn more about the outcome of the attack, the conditional win probability changes based on this information. To formalize this, we introduce four random variables to represent the outcome at each split:
\begin{align*}
    X_i^{(1)} &= \begin{cases}
        1   & \mbox{if attack $i$ results in an attack error}\\
        0   & \mbox{otherwise}
    \end{cases},\\
    X_i^{(2)} &= \begin{cases}
        1   & \mbox{if attack $i$ results in a clean attack (no block touch)}\\
        0   & \mbox{otherwise}
    \end{cases},\\
    X_i^{(3)} &= \begin{cases}
        1   & \mbox{if attack $i$ results in a block error}\\
        0   & \mbox{otherwise}
    \end{cases},\\
    X_i^{(4)} &= \begin{cases}
        1   & \mbox{if attack $i$ results in a block-through}\\
        0   & \mbox{otherwise}
    \end{cases}.
\end{align*}
We use $x_i^{(k)}$ to denote the observed value of the random variable $X_i^{(k)}$. Using these variables, we define seven component models for the outcome of an attack. In each of the models below, the player, team and conference effects are modeled as random effects, each with mean zero and a fixed, unknown variance.\\

\begin{enumerate}
    \item
        {\bf Attack error indicator.} We model the change in conditional point win probability due to whether or not the attack results in an attack error. Our outcome variable $Y_i^{(1)}$ is the difference in win probability before and after $X_i^{(1)}$ is observed:
        \begin{equation*}
            Y_i^{(1)} \equiv \mathbb{E}\left[w(S'_i) \mid S_i, X_i^{(1)}\right] - \mathbb{E}\left[w(S'_i) \mid S_i\right].
        \end{equation*}
        We fit a normal regression model for $Y_i^{(1)}$ with random effects for offensive team, offensive conference, attacker, and setter. The implicit assumption is that the defense does not have a substantial effect on whether the attacker makes an error. We do not include this component model in our strength of schedule estimation, but we do use this model for dividing credit between the attacker and the setter for changes in win probability (Section \ref{sec:attribution-attack}).
        \begin{equation}
        \label{eqn:attack-model-1}
            Y_i^{(1)} \sim \mbox{Normal}\left(
                \alpha + \gamma_{c_i} + \tau_{t_i} + \theta_{a_i} + \psi_{s_i},\,
                \sigma^2_\epsilon
            \right).
        \end{equation}
    \item
        {\bf Clean attack indicator.} We model the change in conditional point win probability due to whether or not the attack results in a clean attack. Our outcome variable $Y_i^{(2)}$ is the difference in win probability before and after $X_i^{(2)}$ is observed:
        \begin{align*}
            Y_i^{(2)} \equiv &~\mathbb{E}\left[w(S'_i) \mid S_i, X_i^{(1)} = 0, X_i^{(2)}\right]\\
            - &~\mathbb{E}\left[w(S'_i) \mid S_i, X_i^{(1)} = 0\right].
        \end{align*}
        We estimate a normal regression model for $Y_i^{(2)}$ with random effects for offensive team, offensive conference, attacker, setter, defensive team, defensive conference, and blocker. We fit this model using only attacks for which $x_i^{(1)} = 0$, i.e. no attack errors. If $x_i^{(2)} = 0$, there is a block touch, and we observe the blocker identity. If $x_i^{(2)} = 1$, there is no block touch, and we infer blocker identity (Section \ref{sec:block-dig-responsibility}).
        \begin{equation}
        \label{eqn:attack-model-2}
            Y_i^{(2)} \sim \mbox{Normal}\left(
                \alpha + \gamma_{c_i} + \tau_{t_i} + \theta_{a_i} + \psi_{s_i} + \tilde\gamma_{\tilde c_i} + \tilde\tau_{\tilde t_i} + \beta_{b_i},\,
                \sigma^2_\epsilon
            \right).
        \end{equation}
    \item
        {\bf Block error indicator.} We model the change in conditional point win probability due to whether or not the attack results in a block error. Our outcome variable $Y_i^{(3)}$ is the difference in win probability before and after $X_i^{(3)}$ is observed:
        \begin{align*}
            Y_i^{(3)} \equiv &~\mathbb{E}\left[w(S'_i) \mid S_i, X_i^{(1)} = 0, X_i^{(2)} = 0, X_i^{(3)}\right]\\
            - &~\mathbb{E}\left[w(S'_i) \mid S_i, X_i^{(1)} = 0, X_i^{(2)} = 0\right].
        \end{align*}
        We estimate a normal regression model for $Y_i^{(3)}$ with random effects for offensive team, offensive conference, attacker, setter, defensive team, defensive conference, and blocker. We fit this model using only attacks for which $x_i^{(1)} = x_i^{(2)} = 0$, i.e. no attack errors, no clean attacks. In this case we always observe the blocker identity. In this case we always observe blocker identity.
        \begin{equation}
        \label{eqn:attack-model-3}
            Y_i^{(3)} \sim \mbox{Normal}\left(
                \alpha + \gamma_{c_i} + \tau_{t_i} + \theta_{a_i} + \psi_{s_i} + \tilde\gamma_{\tilde c_i} + \tilde\tau_{\tilde t_i} + \beta_{b_i},\,
                \sigma^2_\epsilon
            \right).
        \end{equation}
    \item
        {\bf Block-through indicator.} We model the change in conditional point win probability due to whether or not the attack results in a block-through. Our outcome variable $Y_i^{(4)}$ is the difference in win probability before and after $X_i^{(4)}$ is observed:
        \begin{align*}
            Y_i^{(4)} \equiv &~\mathbb{E}\left[w(S'_i) \mid S_i, X_i^{(1)} = 0, X_i^{(2)} = 0, X_i^{(3)} = 0, X_i^{(4)}\right]\\
            - &~\mathbb{E}\left[w(S'_i) \mid S_i, X_i^{(1)} = 0, X_i^{(2)} = 0, X_i^{(3)} = 0\right].
        \end{align*}
        We estimate a normal regression model for $Y_i^{(4)}$ with random effects for offensive team, offensive conference, attacker, setter, defensive team, defensive conference, and blocker. We fit this model using only attacks for which $x_i^{(1)} = x_i^{(2)} = x_i^{(3)} = 0$, i.e. no attack errors, no clean attacks, no block errors. In this case we always observe blocker identity.
        \begin{equation}
        \label{eqn:attack-model-4}
            Y_i^{(4)} \sim \mbox{Normal}\left(
                \alpha + \gamma_{c_i} + \tau_{t_i} + \theta_{a_i} + \psi_{s_i} + \tilde\gamma_{\tilde c_i} + \tilde\tau_{\tilde t_i} + \beta_{b_i},\,
                \sigma^2_\epsilon
            \right).
        \end{equation}
    \item
        {\bf Block-return outcome.} We model the change in point win probability due to the outcome of a block-return. Our outcome variable $Y_i^{(5)}$ is the difference between the win probability of the outcome and the expected win probability, given that a block-return occurs.
        \begin{equation*}
            Y_i^{(5)} \equiv w(S'_i) - \mathbb{E}\left[w(S'_i) \mid S_i, X_i^{(1)} = 0, X_i^{(2)} = 0, X_i^{(3)} = 0, X_i^{(4)} = 1\right].
        \end{equation*}
        We estimate a normal regression model for $Y_i^{(5)}$ with random effects for offensive team, offensive conference, attacker, setter, defensive team, defensive conference, and blocker. We fit this model using only attacks for which $x_i^{(1)} = x_i^{(2)} = x_i^{(3)} = x_i^{(4)} = 0$, i.e. only block-returns. In this case we always observe blocker identity.
        \begin{equation}
        \label{eqn:attack-model-5}
            Y_i^{(5)} \sim \mbox{Normal}\left(
                \alpha + \gamma_{c_i} + \tau_{t_i} + \theta_{a_i} + \psi_{s_i} + \tilde\gamma_{\tilde c_i} + \tilde\tau_{\tilde t_i} + \beta_{b_i},\,
                \sigma^2_\epsilon
            \right).
        \end{equation}
    \item
        {\bf Block-through outcome.} We model the change in point win probability due to the outcome of a block-through. Our outcome variable $Y_i^{(6)}$ is the difference between the win probability of the outcome and the expected win probability, given that a block-through occurs.
        \begin{equation*}
            Y_i^{(6)} \equiv w(S'_i) - \mathbb{E}\left[w(S'_i) \mid S_i, X_i^{(1)} = 0, X_i^{(2)} = 0, X_i^{(3)} = 0, X_i^{(4)} = 0\right].
        \end{equation*}
        We estimate a normal regression model for $Y_i^{(6)}$ with random effects for offensive team, offensive conference, attacker, setter, defensive team, defensive conference, blocker, and digger. We fit this model using only attacks for which $x_i^{(4)} = 1$, i.e. only block-throughs. In this case we always observe blocker identity. We may or may not observe digger identity, depending on whether there is a dig touch after the block touch. If there is no dig touch, we infer digger identity.
        \begin{equation}
        \label{eqn:attack-model-6}
            Y_i^{(6)} \sim \mbox{Normal}\left(
                \alpha + \gamma_{c_i} + \tau_{t_i} + \theta_{a_i} + \psi_{s_i} + \tilde\gamma_{\tilde c_i} + \tilde\tau_{\tilde t_i} + \beta_{b_i} + \delta_{d_i},\,
                \sigma^2_\epsilon
            \right).
        \end{equation}
    \item
        {\bf Clean attack outcome.} We model the change in point win probability due to the outcome of a clean attack. Our outcome variable $Y_i^{(7)}$ is the difference between the win probability of the outcome and the expected win probability, given that a clean attack occurs.
        \begin{equation*}
            Y_i^{(7)} \equiv w(S'_i) - \mathbb{E}\left[w(S'_i) \mid S_i, X_i^{(1)} = 0, X_i^{(2)} = 1, X_i^{(3)} = 0, X_i^{(4)} = 0\right].
        \end{equation*}
        We estimate a normal regression model for $Y_i^{(7)}$ with random effects for offensive team, offensive conference, attacker, setter, defensive team, defensive conference, blocker, and digger. We fit this model using only attacks for which $x_i^{(2)} = 1$, i.e. only clean attacks. In this case there is no block touch, and we must always infer blocker identity. We may or may not observe digger identity, depending on whether there is a dig touch after the block touch. If there is no dig touch, we infer digger identity.
        \begin{equation}
        \label{eqn:attack-model-7}
            Y_i^{(7)} \sim \mbox{Normal}\left(
                \alpha + \gamma_{c_i} + \tau_{t_i} + \theta_{a_i} + \psi_{s_i} + \tilde\gamma_{\tilde c_i} + \tilde\tau_{\tilde t_i} + \beta_{b_i} + \delta_{d_i},\,
                \sigma^2_\epsilon
            \right).
        \end{equation}
\end{enumerate}

In each of the models above, the estimated strength of schedule faced by a player or by a pair of teammates is the sum of the predicted random effects corresponding to the opponents. For example, in model (\ref{eqn:attack-model-7}) the estimated strength of schedule faced by the blocker and the digger is given by $\hat\gamma_{c_i} + \hat\tau_{t_i} + \hat\theta_{a_i} + \hat\psi_{s_i}$. We use $y_i^{(k)}$ to denote the observed value of the random variable $Y_i^{(k)}$, for $k \in \{1, ..., 7\}$.

\subsubsection{Block/Dig Responsibility Assignment}
\label{sec:block-dig-responsibility}

If a block touch occurs following an attack, we observe the identity of the blocker. Otherwise, we must infer it. For models 3 (block error indicator), 4 (block-through indicator), 5 (block-return outcome) and 6 (block-through outcome) in Section \ref{sec:sos-attack}, we always observe blocker identity. For model 2 (clean attack indicator), whether we observe blocker identity depends on the outcome. For model 7 (clean attack outcome), we never observe blocker identity. Although the blocker does not touch the ball, domain knowledge tells us that the blocker plays an important role on clean attack outcomes because the primary responsibility of the blocker is taking away options to channel the attack to the diggers.

For cases in which the blocker identity is unknown, we use a deterministic method for assigning blocker responsibility based on the attack code (i.e. the attacker and type of attack). For each attack code, we identify the defensive position (front left, front middle or front right) which is most frequently responsible for the block touch in the cases when there is a block touch. For cases of this attack code when there is no block touch, we assign the blocker identity to the player in the most frequently responsible position.

Similarly, if a dig touch occurs following an attack, we observe the identity of the digger. Otherwise, we must infer it. In both models 6 and 7, the digger identity may not be observed depending on the outcome. For the unknown digger identities we use a deterministic method for assigning responsibility based on the attack code and the defensive zone (Figure \ref{fig:volleyball-court-diagram}) in which the attack lands. For each combination of attack code and defensive zone, we identify the defensive position (front or back; left, middle or right) which is most frequently responsible for the dig touch in the cases when there is a dig touch. For cases of this attack code and defensive zone when there is no dig touch, we assign the digger identity to the player in the most frequently responsible position.

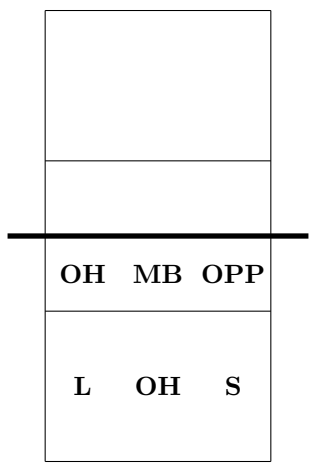
\begin{figure}[hb]
    \centering
    \begin{tikzpicture}
        \draw[line width = 2pt] (-0.5, 3) -- (3.5, 3);
        \draw (0, 0) -- (0, 6);
        \draw (3, 0) -- (3, 6);
        \draw (0, 0) -- (3, 0);
        \draw (0, 2) -- (3, 2);
        \draw (0, 4) -- (3, 4);
        \draw (0, 6) -- (3, 6);
        \node (1) at (2.5, 1) {\bf S};
        \node (2) at (2.5, 2.5) {\bf OPP};
        \node (3) at (1.5, 2.5) {\bf MB};
        \node (4) at (0.5, 2.5) {\bf OH};
        \node (5) at (0.5, 1) {\bf L};
        \node (6) at (1.5, 1) {\bf OH};
    \end{tikzpicture}
    \caption{\it Assumed defensive alignment (when the setter is back row) for blocker and digger identification. When the setter is front row, S and OPP switch places.}
    \label{fig:defensive-alignment}
\end{figure}

Generally, when defending an attack, a team will line up so that one player each is responsible for front-left, front-middle, front-right, back-left, back-middle and back-right. Our data do not tell us how the players are lined up, so we infer this. From the data, we know the rotation spot of the setter. We assume each team sets their service rotation relative to the setter in the following order: S, OH, MB, OPP, OH, MB. Domain knowledge tells us that this is true for the vast majority of teams. Based on the rotation spot of the setter, we know which players start the point in the front row and which players are in the back row (these do not change during the point). We assume that players line up defensively according to Figure \ref{fig:defensive-alignment}. Again, domain knowledge tells us that most (not all) teams used this alignment.

Note that in Figure \ref{fig:defensive-alignment} we have assumed that the libero has replaced whichever middle blocker is in the back row. Our data do not indicate when the libero comes on and off the court or even which player is the libero. We infer the identity of the libero in each set using the fact that she is the only player who makes contacts without appearing in the lineup.

\subsection{Individual Attribution}
\label{sec:attribution}

\subsubsection{Serve/Receive Attribution}

Credit attribution for the serve/receive outcome is straightforward because there is one player on each side of the adversarial matchup. For each player, we credit her with the difference between the change in point win probability and her strength of schedule. On serve $i$, we define the Points Gained (PG) by the serving player and receiving player, respectively, as
\begin{align}
    \label{eqn:pg-serve}
    \begin{split}
        \mbox{PG}_i^{\mbox{\scriptsize SV}} &= y_i - \left(\hat\alpha + \hat{\tilde\gamma}_{c_i} + \hat{\tilde\tau}_{t_i} + \hat\pi_{p_i}\right),\\
        \mbox{PG}_i^{\mbox{\scriptsize R}} &= -\left(y_i - \left(\hat\alpha + \hat\gamma_{c_i} + \hat\tau_{t_i} + \hat\pi_{p_i}\right)\right).
    \end{split}
\end{align}

Importantly, the unit of Points Gained is points. This allows us to make apples-to-apples comparisons across different skills of individual contributions to team success. We measure these contributions in terms of point differential, which is highly interpretable and connects directly to winning matches.

\subsubsection{Set/Attack/Block/Dig Attribution}
\label{sec:attribution-attack}

Credit attribution for the attack/block/dig outcome is more sophisticated because there are two players on each side of the adversarial matchup and because we have seven component models with different random effect specifications. We first perform the credit attribution within each of the component models separately, and then we sum the credit across components. To divide credit between teammates, we turn to the variance components estimated in the random-effects linear regression models (\ref{eqn:attack-model-1}) through (\ref{eqn:attack-model-7}).

Taking model (\ref{eqn:attack-model-1}) for example, let $\sigma^2_\theta$ be the variance of the random effects $\theta_a$ for $a \in \mathcal{P}$, and let $\sigma^2_\psi$ be the variance of the random effects $\psi_s$ for $s \in \mathcal P$. The attacker identity explains $\sigma^2_\theta / (\sigma^2_\theta + \sigma^2_\psi)$ of the variance in the contribution of attacker and setter identities to the expected outcome. In other words, the fraction of variance in $\theta_{a_i} + \psi_{s_i}$ explained by the attacker identity is $\sigma^2_\theta / (\sigma^2_\theta + \sigma^2_\psi)$, and the fraction of variance explained by the setter identity is $\sigma^2_\psi / (\sigma^2_\theta + \sigma^2_\psi)$. We use the estimate of this ratio divide credit for $y_i^{(1)}$ between the attacker and the setter. For attack $i$, we define the Points Gained on component (1) by the attacker and setter, respectively, as:
\begin{align}
    \label{eqn:pg-attack-1}
    \begin{split}
        \mbox{PG}_i^{\mbox{\scriptsize A(1)}} &= \frac{\hat\sigma^2_\theta}{(\hat\sigma^2_\theta + \hat\sigma^2_\psi)} \cdot y_i^{(1)},\\
        \mbox{PG}_i^{\mbox{\scriptsize S(1)}} &= \frac{\hat\sigma^2_\psi}{(\hat\sigma^2_\theta + \hat\sigma^2_\psi)} \cdot y_i^{(1)}.
    \end{split}
\end{align}

Model (\ref{eqn:attack-model-1}) does not include a strength of schedule adjustment for the attacker and setter, but consider model (\ref{eqn:attack-model-2}). For attack $i$, if $x_i^{(1)} = 1$ (attack error), then the Points Gained on component (2) are zero for all players. Otherwise we define the Points Gained on component (2) by the attacker, setter and blocker, respectively, as:
\begin{align}
    \label{eqn:pg-attack-2}
    \begin{split}
        \mbox{PG}_i^{\mbox{\scriptsize A(2)}} &= \frac{\hat\sigma^2_\theta}{(\hat\sigma^2_\theta + \hat\sigma^2_\psi)} \cdot \left(y_i^{(2)} - (\hat\alpha + \hat{\tilde\gamma}_{\tilde c_i} + \hat{\tilde\tau}_{\tilde t_i} + \hat\beta_{b_i})\right),\\
        \mbox{PG}_i^{\mbox{\scriptsize S(2)}} &= \frac{\hat\sigma^2_\psi}{(\hat\sigma^2_\theta + \hat\sigma^2_\psi)} \cdot \left(y_i^{(2)} - (\hat\alpha + \hat{\tilde\gamma}_{\tilde c_i} + \hat{\tilde\tau}_{\tilde t_i} + \hat\beta_{b_i})\right),\\
        \mbox{PG}_i^{\mbox{\scriptsize B(2)}} &= -\left(y_i^{(2)} - (\hat\alpha + \hat\gamma_{c_i} + \hat\tau_{t_i} + \hat\theta_{a_i} + \hat\psi_{s_i})\right).
    \end{split}
\end{align}
Because the blocker does not have a teammate represented in this component model, there is no splitting of credit based on variance components. We do not enumerate the definition of Points Gained for all of the attack outcome models because the extension is straightforward. We provide one final example for model (\ref{eqn:attack-model-7}). Let $\sigma^2_\beta$ be the variance of the random effects $\beta_b$ for $b \in \mathcal{P}$, and let $\sigma^2_\delta$ be the variance of the random effects $\delta_d$ for $d \in \mathcal P$. For attack $i$, if $x_i^{(2)} = 0$ (no clean attack), then the Points Gained on component (7) are zero for all players. Otherwise we define the Points Gained on component (7) by the attacker, setter, blocker and digger, respectively, as:
\begin{align}
    \label{eqn:pg-attack-7}
    \begin{split}
        \mbox{PG}_i^{\mbox{\scriptsize A(7)}} &= \frac{\hat\sigma^2_\theta}{(\hat\sigma^2_\theta + \hat\sigma^2_\psi)} \cdot \left(y_i^{(7)} - (\hat\alpha + \hat{\tilde\gamma}_{\tilde c_i} + \hat{\tilde\tau}_{\tilde t_i} + \hat\beta_{b_i} + \hat\delta_{d_i})\right),\\
        \mbox{PG}_i^{\mbox{\scriptsize S(7)}} &= \frac{\hat\sigma^2_\psi}{(\hat\sigma^2_\theta + \hat\sigma^2_\psi)} \cdot \left(y_i^{(7)} - (\hat\alpha + \hat{\tilde\gamma}_{\tilde c_i} + \hat{\tilde\tau}_{\tilde t_i} + \hat\beta_{b_i} + \hat\delta_{d_i})\right),\\
        \mbox{PG}_i^{\mbox{\scriptsize B(7)}} &= -\frac{\hat\sigma^2_\beta}{(\hat\sigma^2_\beta + \hat\sigma^2_\delta)} \cdot \left(y_i^{(7)} - (\hat\alpha + \hat\gamma_{c_i} + \hat\tau_{t_i} + \hat\theta_{a_i} + \hat\psi_{s_i})\right),\\
        \mbox{PG}_i^{\mbox{\scriptsize D(7)}} &= -\frac{\hat\sigma^2_\delta}{(\hat\sigma^2_\beta + \hat\sigma^2_\delta)} \cdot \left(y_i^{(7)} - (\hat\alpha + \hat\gamma_{c_i} + \hat\tau_{t_i} + \hat\theta_{a_i} + \hat\psi_{s_i})\right).
    \end{split}
\end{align}

We sum Points Gained across all components to obtain the Points Gained by each player on each attack:
\begin{equation}
    \label{eqn:pg-attack}
    \mbox{PG}_i^{\mbox{\scriptsize A}} = \sum_{k = 1}^7\mbox{PG}_i^{\mbox{\scriptsize A(k)}}, \hspace{5mm}
    \mbox{PG}_i^{\mbox{\scriptsize S}} = \sum_{k = 1}^7\mbox{PG}_i^{\mbox{\scriptsize S(k)}}, \hspace{5mm}
    \mbox{PG}_i^{\mbox{\scriptsize B}} = \sum_{k = 2}^7\mbox{PG}_i^{\mbox{\scriptsize B(k)}}, \hspace{5mm}
    \mbox{PG}_i^{\mbox{\scriptsize D}} = \sum_{k = 6}^7\mbox{PG}_i^{\mbox{\scriptsize D(k)}}.
\end{equation}

\section{Results}

\subsection{Strength of Schedule}

\begin{figure}
    \centering
    \includegraphics[width=0.5\textwidth]{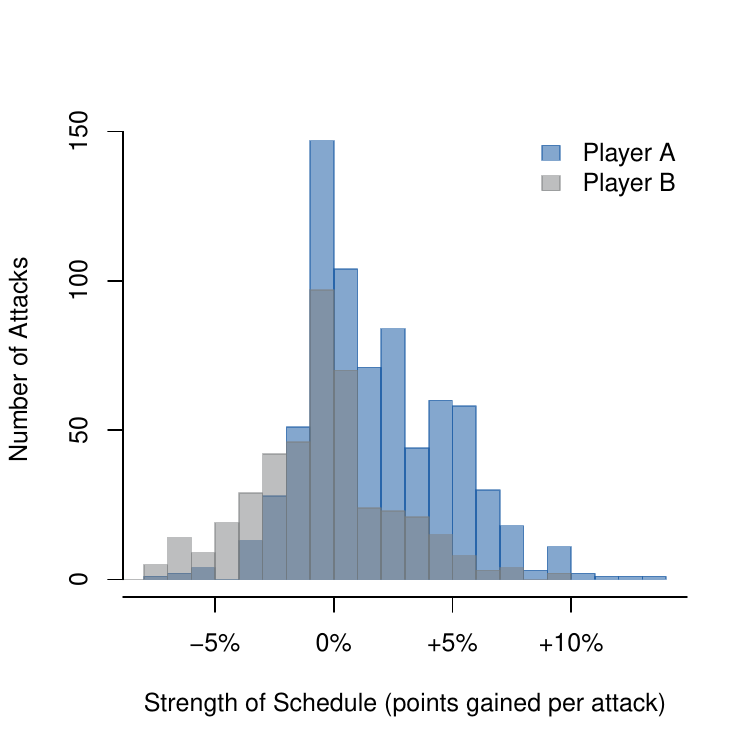}
    \caption{\it Distributions of strength of schedule (points per attack) for two teammates. Player A and Player B are teammates on a team in a Power 5 conference. They are the two primary outside hitters for their team. For each attack by either player, we have an SoS estimate that can be interpreted as the reduction in point win probability attributable to quality of competition (higher SoS = tougher competition). The blue histogram shows the SoS distribution for Player A, and the gray histogram shows the SoS distribution for Player B.}
    \label{fig:teammate-comparison}
\end{figure}

We start with a description of the results from estimating the strength of schedule (SoS) models described in Section \ref{sec:strength-of-schedule}. Note that SoS is defined on a contact-by-contact level, based on the specific players matched up against each other on each adversarial contact. This means that teammates can have different strengths of schedule. Consider, for example, the two teammates compared in Figure \ref{fig:teammate-comparison}. These teammates were the two primary outside hitters for a team in a major conference. We observe that Player A faced a much tougher SoS (+1.4\% PG per attack) than Player B (--0.1\% PG per attack). The single toughest SoS faced on an attack by either player was +13.0\% when Player A recorded an attack against blocker Kaitlyn Hord and digger Lexi Rodriguez of Nebraska.

We can aggregate SoS on a player, team or conference level by calculating the average SoS within the desired group. For example, Table \ref{tab:top-ten-conferences} shows the top ten average strength of schedule for Points Gained per set (aggregated across all skills) on a conference level. We observe that the Power 5 conferences are the top conferences by average SoS, as we would hope to see from a good strength of schedule model.

\begin{table}
    \centering
      \begin{tabular}{c|c}
    Conference      & Avg SoS\\
    \hline
    Big Ten         & +0.23\\
    Pac-12          & +0.23\\
    SEC             & +0.21\\
    Big 12          & +0.20\\
    ACC             & +0.15\\
    West Coast      & +0.09\\
    American        & +0.04\\
    Big West        & +0.04\\
    Mountain West   & +0.02\\
    Mid-American    & +0.01\\
  \end{tabular}
    \caption{\it Top ten conferences by average strength of schedule (points per set). SoS is defined on a contact-by-contact level, and we aggregate it at the conference level by averaging. These SoS numbers are to be interpreted as the average team-wide reduction in point differential per set based on the quality of competition faced by teams in the conference.}
    \label{tab:top-ten-conferences}
\end{table}

\begin{figure}
    \centering
    \includegraphics[width=\textwidth]{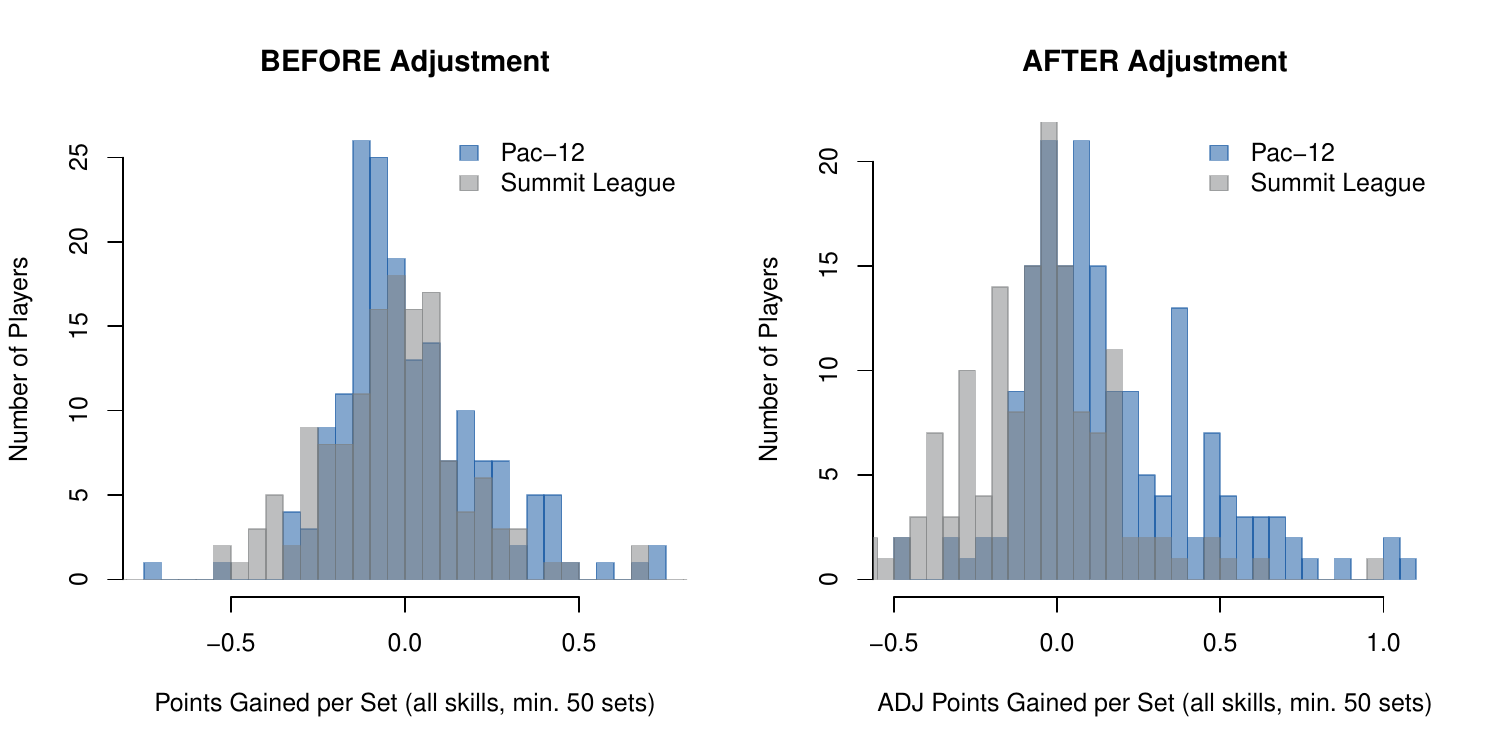}
    \caption{\it Comparison of Pac-12 and Summit League player performances before and after strength of schedule adjustment. Points Gained per set is interpretable as the contribution to team point differential by each player, relative to an average player. The blue histogram shows the distribution of player performance in the Pac-12 while the gray histogram shows the distribution of player performance in the Summit League.}
    \label{fig:conference-comparison}
\end{figure}

Figure \ref{fig:conference-comparison} illustrates the value of the strength of schedule adjustments for player evaluation. We show distributions of player Points Gained per set between the two conferences before ({\it raw} Points Gained) and after ({\it adjusted} Points Gained) the SoS adjustment. We compare the Pac-12, one of the top conferences in the NCAA, with the Summit League, which is closer to the middle of our conference SoS rankings. Before SoS adjustment, there is not much separation between the performance metrics for players in the two conferences. This makes sense because performance is measured relative to competition. After the SoS adjustment, we see more separation between the two distributions. Between the two conferences, most of the top players by adjusted Points Gained per set are in the Pac-12 (although one player from the Summit League stands out---she will stand out again in Section \ref{sec:points-gained}).

\subsection{Individual Contributions to Team Success}
\label{sec:points-gained}

\begin{figure}
    \centering
    \includegraphics[width=\textwidth]{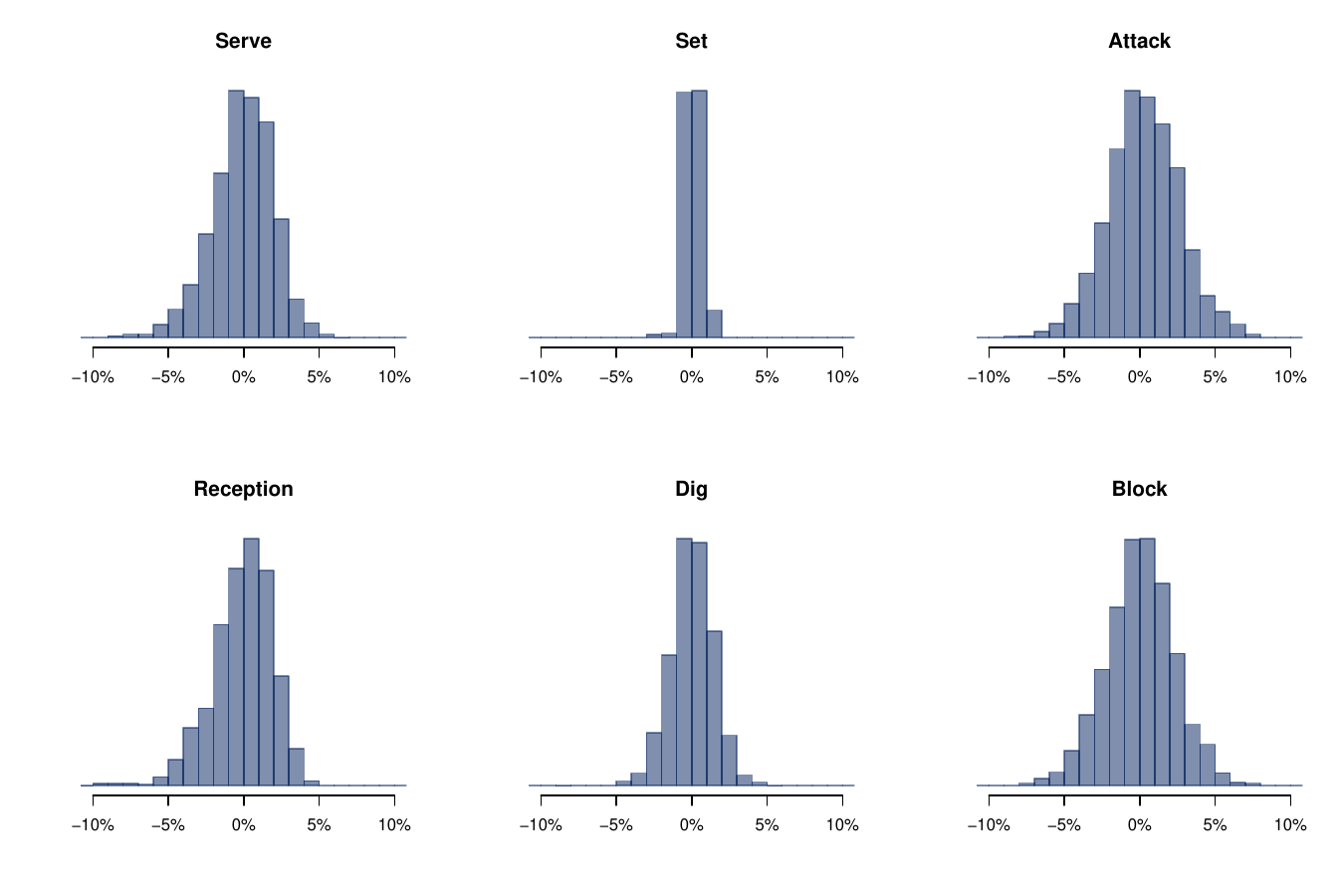}
    \caption{\it Distribution of player Points Gained per contact for different skills. Points Gained per contact is interpretable as how much the player increases her team's probability of winning the point every time she performs the skill. This histograms show the distribution of players by average Points Gained per contact. The minimum sample sizes for serve, set, attack, reception, dig, block are 100, 1000, 200, 100, 100, 200 contacts, respectively.}
    \label{fig:points-gained-per-opportunity}
\end{figure}

Figure \ref{fig:points-gained-per-opportunity} shows the distribution of raw Points Gained per contact, for each type of contact (serve, reception, set, attack, dig and block). This result highlights the power of the point win probability model and the approach of measuring player contributions based on changes in point win probability. Because the unit of measurement is the same across skills (points), we can make comparisons across skills and evaluate trade-offs between skills. Somewhat surprisingly, the magnitude of the effect of each skill on point win probability is roughly similar on a per-contact basis, as measured by the spread of each histogram in Figure \ref{fig:points-gained-per-opportunity}. The one exception is setting, for which the per-contact spread of raw Points Gained per opportunity is significantly less than the same spread for the other skills. However, note that set contacts are generally concentrated to one player (if the team is running a 5-1 scheme, or two if the team is running a 6-2 scheme) whereas attack contacts are distributed across five to six players, so the setter can impact the game at a level comparable to her attackers. Set and attack are the most frequent types of contacts ($\sim23\%$ each), followed by dig, serve and reception ($\sim15\%$ each). The least frequently recorded contact are blocks ($\sim9\%$).

The fact that the spread of raw Points Gained per contact is so much smaller for setting than for attacking is is due to the ratio of $\hat\sigma^2_\psi / (\hat\sigma^2_\theta + \hat\sigma^2_\psi)$ as in equations (\ref{eqn:pg-attack-1})--(\ref{eqn:pg-attack-7}). The model has learned from the data that attacker identity explains more variation in attack outcome than does setter identity. Table \ref{tab:outcome-tree-percents} reports the attribution percentage for each pair of teammates for each node in the attack outcome tree. The component outcomes for which the setter gets the most attribution are (2) whether there is a clean attack or a block touch and (7) the result of a clean attack. This result aligns well with domain knowledge because the setter is responsible for choosing whom to set and deceiving the opposing blockers, making a block touch less likely. Interestingly, we have learned through our modeling that the digger identity explains (relative to blocker identity) less of the variance in (6) the result of a block-through than in (7) the result of a clean attack. In other words, when the blocker gets a touch on the ball, the digger matters more than the blocker (relative to when the blocker does not get a touch on the ball).

\begin{table}
    \centering
    \begin{tabular}{c|ccccccc}
    &  (1)&  (2)&  (3)&  (4)&  (5)&  (6)&  (7)\\
    \hline Attacker&  91\%&  70\%&  81\%&  73\%&  89\%&  79\%&  69\%\\
    Setter&  9\%&  30\%&  19\%&  27\%&  11\%&  21\%&  31\%\\ \hline Blocker& --& 100\%& 100\%& 100\%& 100\%& 28\%& 52\%\\
    Digger&  --&  --&  --&  --&  --&  72\%&  48\%\\
\end{tabular}

    \caption{\it Division of Points Gained between teammates for each split of the attack outcome tree. The parenthetical column labels correspond to the the parenthetical split labels in Figure \ref{fig:attack-model-tree}. At each split, the change in conditional point win probability before and after the split is shared between the teammates involved, according to the percentages in this table.}
    \label{tab:outcome-tree-percents}
\end{table}

Outside hitters (OHs) generally have the highest volume of attacks, and they also have more opportunities than other attackers to impact the game through receptions and digs. So it is no surprise that the ranking of top players by adjusted Points Gained per set played is dominated by outside hitters. Table \ref{tab:top-ten-players} shows the top ten players across the NCAA in 2022 by adjusted Points Gained per set, along with a breakdown of the skills that constitute their contributions. For all ten players (except Asjia O'Neil, who contributes most through blocking), the biggest contribution comes from their attacking. All of the players on this list received at least All-America honorable mention, including five first-team All-Americans \citep{avca_all_america}. A notable case is Elizabeth Juhnke (All-America honorable mention), a standout performer in the Summit League (not one of the top 10 leagues in Table \ref{tab:top-ten-conferences}). Even after adjusting for strength of schedule, Juhnke is still among the top performers in the NCAA. The question of how performance in a mid-major conference will translate to performance in a top conference is an important one for teams which actively recruit players from the NCAA transfer portal.

\begin{table}
    \centering
    \begin{tabular}{ccccc|c|ccccc}
     PLAYER & TEAM&  CONF &  POS &  SETS &  PG*/S&  SRV &  PASS &  SET &ATT &BLK\\
     \hline
     Brooke Nuneviller&  Oregon&  Pac-12&  OH&  122&  +1.09&  +0.07&  +0.41&  +0.00&  +0.56&+0.04\\
     McKenna Melville&  Central Florida&  AAC&  OH&  104&  +1.09&  -0.14&  +0.23&  -0.00&  +0.79&+0.22\\
     Claire Hoffman&  Washington&  Pac-12&  OH&  112&  +1.04&  +0.13&  +0.23&  -0.00&  +0.65&+0.02\\
     Julia Bergmann&  Georgia Tech&  ACC&  OH&  86&  +1.03&  +0.09&  +0.25&  -0.01&  +0.64&+0.06\\
     Kendall Kipp&  Stanford&  Pac-12&  OPP&  117&  +1.02&  +0.03&  -0.02&  -0.00&  +0.72&+0.29\\
     Amber Igiede&  Hawaii&  Big West&  MB&  102&  +0.98&  +0.07&  +0.04&  +0.01&  +0.47&+0.38\\
     Elizabeth Juhnke&  South Dakota&  Summit&  OH&  113&  +0.96&  +0.01&  -0.01&  -0.00&  +0.69&+0.26\\
     Madi Kubik&  Nebraska&  Big Ten&  OH&  109&  +0.94&  +0.05&  +0.42&  -0.01&  +0.44&+0.05\\
     Asjia O'Neal&  Texas&  Big 12&  MB&  87&  +0.93&  +0.05&  +0.04&  +0.00&  +0.35&+0.50\\
Logan Eggleston& Texas& Big 12& OH& 91& +0.89& +0.09& +0.05& +0.01& +0.70&+0.05\\
\end{tabular}

    \caption{\it Top ten players by adjusted Points Gained per set (PG*/S) for the 2022 NCAA season. Here, SETS means sets played (as opposed to the skill of setting). We break down PG*/S into contributions from individual skills: serve, pass (reception and dig), set, attack and block.}
    \label{tab:top-ten-players}
\end{table}

\begin{figure}
    \centering
    \includegraphics[height=5.4cm]{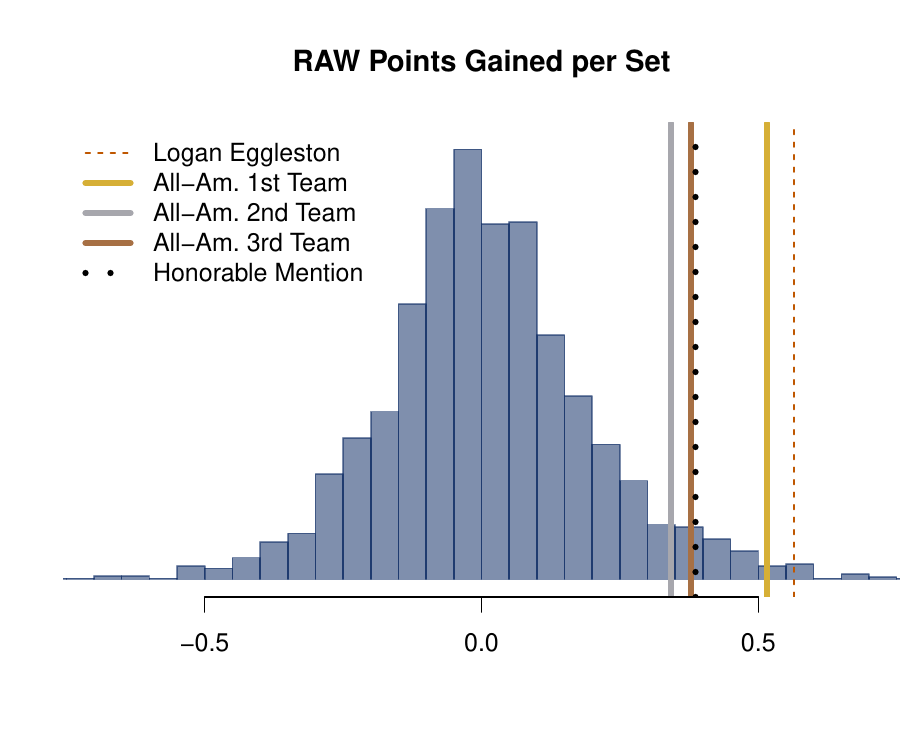}
    \includegraphics[height=5.4cm]{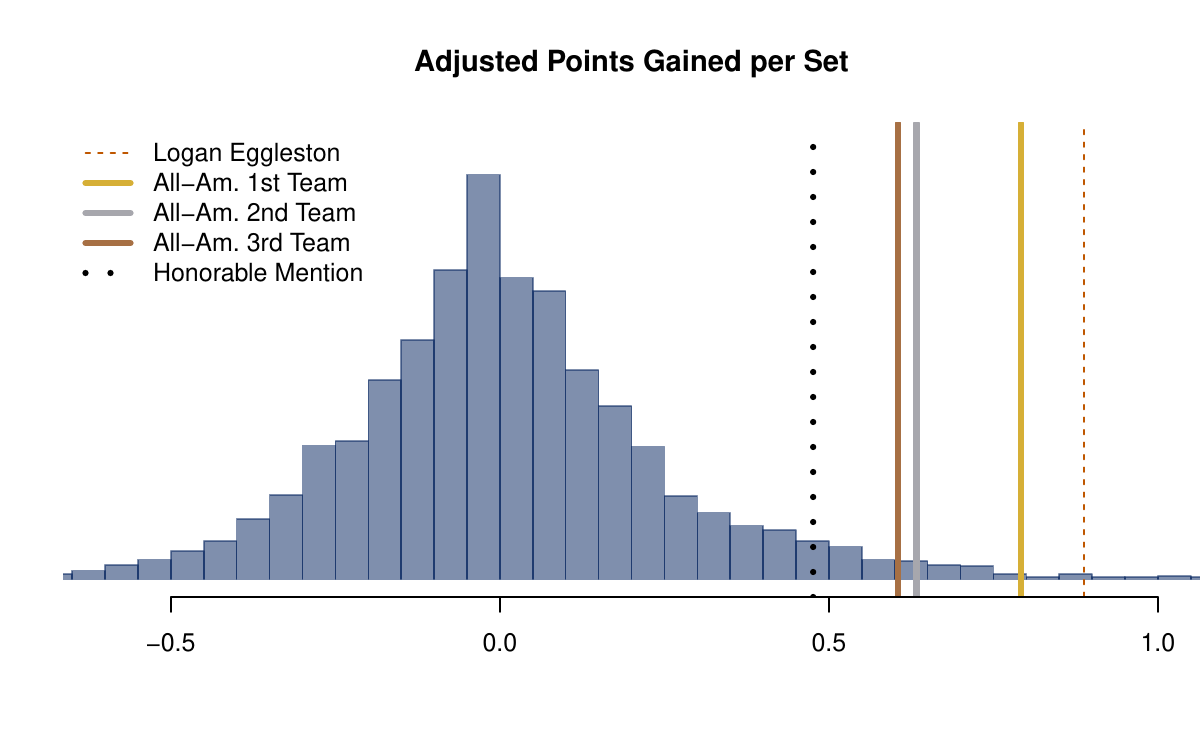}
    \caption{\it Distributions of raw and adjusted Points Gained per set for all NCAA players (minimum 50 sets played). Vertical lines indicate the average PG per set within each category of All-America recognition (1st, 2nd, 3rd, HM). We also indicate national player of the year Logan Eggleston with a vertical line.}
    \label{fig:avca-all-americans}
\end{figure}

Figure \ref{fig:avca-all-americans} shows the observed distribution of Points Gained per set across all players in the NCAA (minimum 50 sets played), both raw and adjusted. This shows exhibits good agreement between Points Gained and subjective evaluations of players by coaches in that the players recognized by All-America awards are among the top performers by Points Gained per set. Satisfyingly, after applying the strength of schedule adjustment, we observe that the ordering of average performance within the recognition categories matches the ordering of the recognition categories.

\begin{figure}
    \centering
    \includegraphics[width=0.5\textwidth]{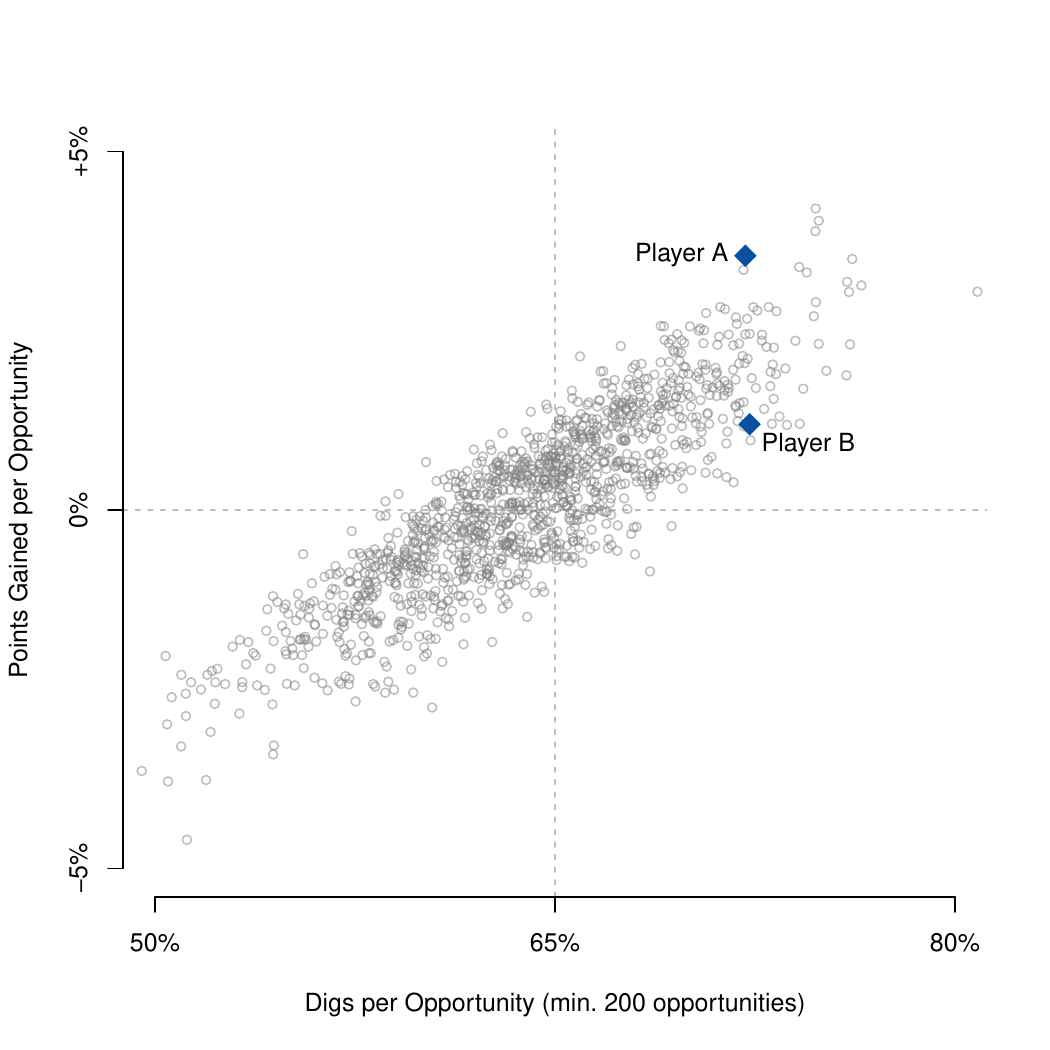}
    \caption{\it Dig Points Gained per opportunity vs. digs per opportunity for all NCAA players (min 200 dig opportunities). Player A and Player B have the same digs per opportunity (72\%), but Player A (+3.6\%) rates much higher by PG per opportunity than Player B (+1.2\%) because of the additional information in Points Gained.}
    \label{fig:dig-comparison}
\end{figure}

To highlight the value of raw Points Gained in contrast with the best player performance metrics readily available to coaches, consider the task of evaluating back-row defensive performance. Traditional box scores offer digs per set although coaches generally have access to digs per opportunity (an opportunity is an attack which the digger is responsible to defend). Figure \ref{fig:dig-comparison} compares digs per opportunity with dig Points Gained per opportunity. Although there is a strong correlation between the two metrics (as expected), Points Gained can reveal differences in performance that are hidden by digs per opportunity. For example, Player A is a libero, and Player B is an outside hitter, each with 72\% digs per opportunity on a large number of opportunities. However, Player A's opportunities were more difficult (85\% were clean attacks) than those for Player B (66\% were clean attacks). And while all digs count the same in the traditional metrics, Player A had a higher rate (48\%) of ``perfect''-rated digs than Player B (36\%). These differences don't show up in digs per opportunity, but they do show up in raw Points Gained. Player A increased her team's point win probability by 3.6\% per dig opportunity (an elite performance) while Player B increased her team's point win probability by 1.2\% per dig opportunity (a good performance).

\section{Applications}
\label{sec:applications}

The obvious application of Points Gained is for player recruitment. There are several thousand women's Division I college volleyball players in the NCAA. College teams and new professional teams in the United States do not have the resources to scout all of these players. By synthesizing information from more than 5 million contacts per season, Points Gained can help teams narrow their focus for scouting. Thanks to the strength of schedule adjustment, these results can also identify players from smaller conferences who may be strong candidates to play on a bigger stage.

Aside from the scouting angle, we also want to illustrate how these results might be applied to in-game decision-making. A common strategy in volleyball is to use a substitute for one of their outside hitters when she is in the back row. This substitute is a {\it defensive specialist} (DS) who specializes in passing (receptions and digs). Across the NCAA, 63\% of teams generally replace at least one of their outside hitters with a DS in the back row. However, for every complete rotation of the team, this strategy costs two substitutions (one to sub the DS in, one to sub the DS out), and substitutions are capped. So how much value is the team deriving from this use of substitutions? And how does this value compare to other potential uses of the substitution (e.g. serving specialist)?

\begin{figure}
    \centering
    \includegraphics[height=5.4cm]{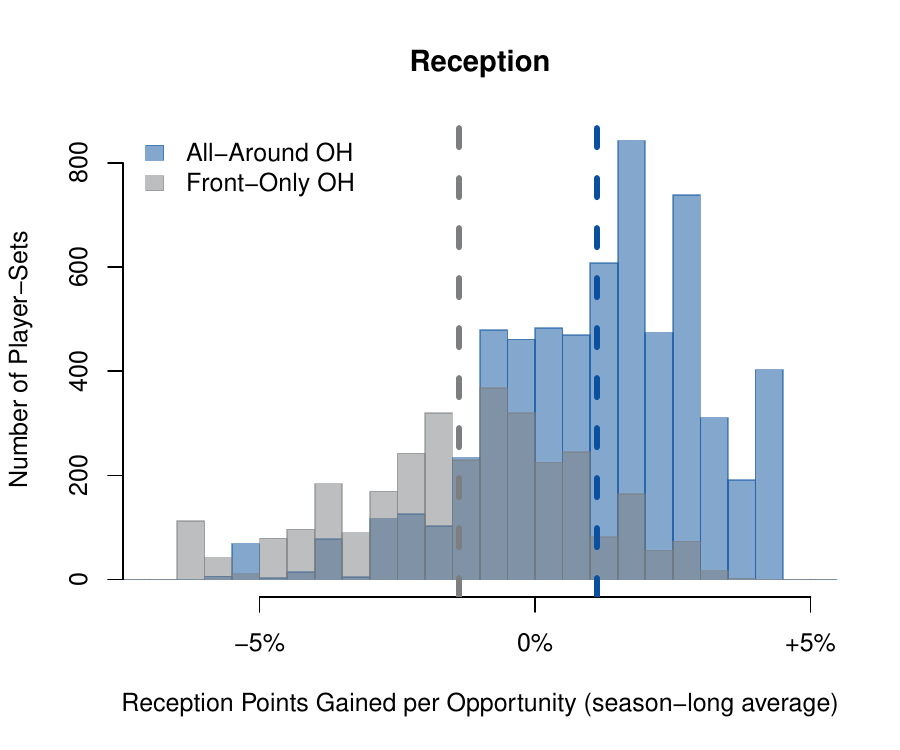}
    \includegraphics[height=5.4cm]{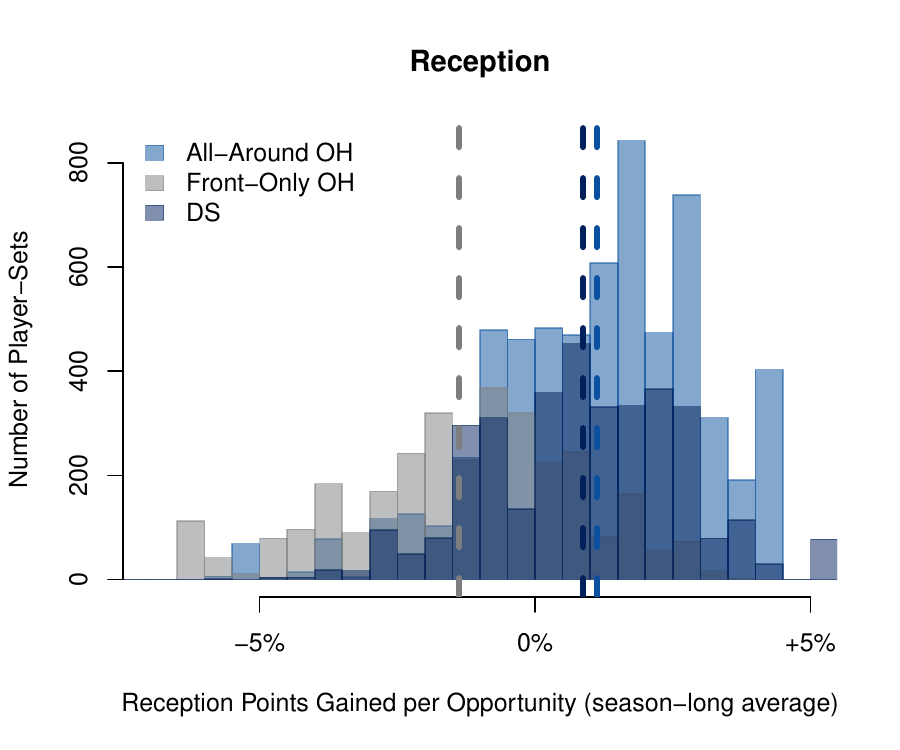}
    \caption{\it Reception Points Gained per opportunity by outside hitters and defensive specialists. The light blue histogram shows the distribution of observed reception performance for outside hitters who are usually not subbed out for defensive specialists (average: +1.1\%). The right blue histogram shows the distribution of observed performance for outside hitters who usually are subbed out for defensive specialist (average: --1.4\%). On the right side, we have overlayed in dark blue the distribution of observed reception performance for defensive specialists (average: +0.9\%).}
    \label{fig:oh-comparison}
\end{figure}

We will not answer this question fully, but we will address one component of it. In the back row, the defensive specialist will perform differently from the outside hitter in three skills: reception, dig and (back-row) attack. Let us focus specifically on the reception skill. How much better is the typical DS at receiving serves, relative to the typical outside hitter? Unsurprisingly, there is a big difference between reception performance among all-around outside hitters who usually play in the back row and reception performance by front-only outside hitters who usually do not play in the back row, as show in Figure \ref{fig:oh-comparison}. Note that outside hitters are generally responsible for serve receive when they are in the front row, which mistakes the problem of estimating this skill for front-only outside hitters.

We see in Figure \ref{fig:oh-comparison} that the distribution of DS reception performance is similar to that of all-around outside hitters. The difference between reception Points Gained per opportunity for defensive specialists and front-only outside hitters is 2.3\%, meaning that every reception opportunity given to the DS increases the team's probability of winning the point by 2.3\%. It turns out that the expected number of substitutable reception opportunities is 0.1 per point (because they only happen when the outside hitter is in the back row, and the serve still needs to go to the DS). So adopting this substitution strategy would increase the team's average point winning percentage by 0.2\%.

\begin{figure}
    \centering
    \includegraphics[height=5.4cm]{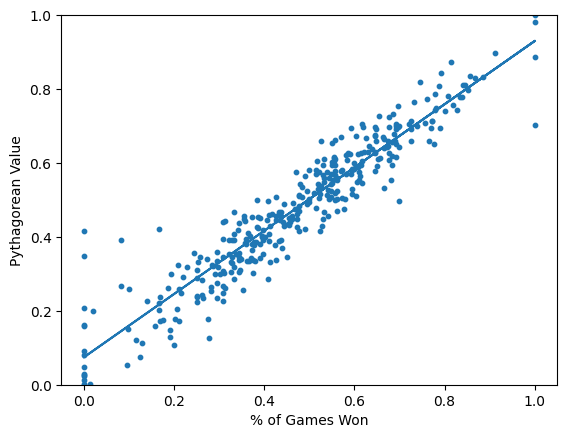}
    \caption{\it Observed winning percentage vs. Pythagorean winning percentage for 2022 NCAA teams. We observe that point ratio is a strong predictor of winning percentage in women's college volleyball.}
    \label{fig:pythag-games-won}
\end{figure}

To analyze the decision fully, we would go through the same exercise for digging and (back-row) attacking. An increase of 0.23\% in point winning percentage sounds very small, but we can put it in context using the Pythagorean formula to connect winning points to winning games. Following \citet{winston_etal_2022}, we define Pythagorean winning percentage as $\mbox{PS}^\alpha / (\mbox{PS}^\alpha + \mbox{PA}^{\alpha})$, where PS is points scored; PA is points allowed; and $\alpha > 0$ is a constant to be estimated. For our data, the optimal $\alpha$ for predicting winning percentage is 9.3. The correlation between Pythagorean winning percentage and actual winning percentage is strong, as illustrated in Figure \ref{fig:pythag-games-won}.

Based on this formula for Pythagorean winning percentage, a team that wins 50\% of its points would win 50\% of its games, and a team that wins 50.2\% of its points would win 52\% of its games. So the effect of replacing the average front-only outside hitter with a defensive specialist is not so small after all. More careful analysis is necessary, but this example illustrates how Points Gained might help coaches answer this strategic question.

\section{Discussion}

We have presented a framework for evaluating volleyball player performance using charted contact-by-contact data. We measure the core skills (serve, reception, set, attack, block and dig) all with the common measurement unit of points, facilitating comparisons between skills. Further, we introduce a sophisticated method for adjust player performance relative to the quality of competition faced by the player on a contact-by-contact basis. Taken together, adjusted Points Gained is a tool that can inform team decision-making on player recruitment by synthesizing information from millions of contacts per season. For a college team active on the transfer portal or for a new professional volleyball team, it presents a starting point for narrowing in on players to scout more carefully.

In addition to the player evaluation application, Points Gained can also form the basis for in-game strategy decisions, as exemplified in Section \ref{sec:applications}. The point win probability model is a fundamental model that facilitates the measurement of decisions in terms of points. In-game decision-making generally involves trade-offs. By measuring disparate contributions to winning on the same scale, we can start to answer these questions with some level of objectivity.

\subsection{Limitations}
\label{sec:limitations}

We acknowledge the following limitations of Points Gained as defined in the present work.

First, there is a bias against setters in dig evaluation codes. The \# evaluation code for a dig means that the digger made a perfect pass to the setter. When the setter digs an attack, the best possible evaluation code is +. This lowers our estimation of setter performance in back row defense relative to other positions.

Second, the deterministic assignment of blocker and digger responsibilities is a coarse simplification of defensive responsibility. For some attack codes, there is close to a 50-50 split in block touch frequency between two different blocker positions. A probabilistic assignment of blocker and digger responsibility. Better still would be a model that acknowledges the possibility of multiple blockers.

Third, the strength of schedule adjustment is outcome-dependent. For example, if an attacker hits the ball at a strong defender, she is credited by adjusted Points Gained for having faced a tough strength of schedule. In reality, hitting the ball at a weaker defender is good performance (not just weak SoS), and a better SoS model would be outcome-agnostic.

Fourth, Points Gained does not leverage the spatial information available in the dataset (e.g. pass end locations), which would lead to more precise estimation of point win probability and individual contributions to team success.

Despite these limitations, Points Gained presents a significant step forward in performance evaluation relative to the industry standard in college volleyball.

\bibliography{arxiv}

\end{document}